# Unveiling Novelty Evolution in the field of Library and Information Science in China

Chen Yang
Nanjing University of Science and Technology, Nanjing, China
Yuzhuo Wang
Anhui University, Hefei, China
Chengzhi Zhang[*]
Nanjing University of Science and Technology, Nanjing, China

**Abstract**

**Purpose -** This study aims to analyze the distribution of novelty among scholarly papers in the field of Library and Information Science (LIS) in China. Specifically, we explore the distribution of novelty of papers in various journals, research topics, and different periods. It is possible to understand the characteristics of library and information science research in China and what factors have influenced it.

**Design/methodology/approach -** This paper collects articles published in Chinese library science journals indexed by the Chinese Social Sciences Citation Index (CSSCI) from 2000 to 2022. The BERTopic model is used based on abstracts of the papers and to obtain the topic of each paper. Based on the combination innovation theory of reference pairs cited by focal papers, novelty scores of all papers are calculated. Next, we analyze the novelty of papers under different topics. Finally, we analyze the differences in author collaboration patterns across various topics, aiming to explain how these differences relate to the novelty of papers from a collaborative perspective.

**Findings -** This study shows that archival research topics have lower novelty than papers on journal evaluation and patent technology in Chinese Library and Information Science. Research papers in this field are gradually becoming more novel over time. Papers on different topics and with varying degrees of novelty exhibit distinct author collaboration patterns, with low-novelty topics more frequently featuring solo authorship, while high-novelty topics tend to involve a higher percentage of inter-institutional collaboration.

**Originality/value -** This study investigates the novelty characteristics of research papers on different topics in the field of LIS in China. Our contribution includes visualizing research hotspots and trends in the field and analyzing authors' collaboration patterns at the level of research topics, thereby providing new perspectives on the factors affecting the novelty of these papers.

[*] Corresponding author, E-mail address: zhangcz@njust.edu.cn (C. Zhang).

# 1. Introduction

In the current era of information explosion, the field of library and information science (LIS), serving as the core domain for information management and dissemination, undertakes a crucial mission (Tuomaala et al., 2014). With the rapid advancement of information technology and the continuous evolution of societal information needs, researchers in this field are increasingly exploring various topics, ranging from information retrieval and knowledge organization to digital libraries and information services(Wang et al., 2023). Different topics represent distinct research directions and issues. Detailed topic analysis makes it possible to unveil the research frontiers, core hotspots, and evolving relationships over time within the LIS field (Ma and Lund, 2020), which aids the academic community in gaining a better understanding of the dynamics within the field, guiding future research directions.

Innovation permeates every aspect of human societal production and practical activities (Zeng et al.,2023). It serves as an inexhaustible driving force for the development of disciplines, and academic research innovation greatly promotes the advancement of disciplinary fields. By examining the novelty of academic papers across different topics in the library and information science domain, new research directions and theoretical viewpoints can be discovered, thereby propelling the academic development of the field (Cao et al.,2020). Moreover, in today's fiercely competitive academic environment, researchers must continuously pursue innovation and breakthroughs to enhance their academic and research capabilities. Through studying the novelty of academic papers across different topics in the library and information science field, researchers can better grasp the academic frontier and strengthen their academic competitiveness.

Scientific research collaboration is a scientific activity in which two or more researchers or organizations work together on the same research task and achieve the goal of maximizing scientific research output through mutual collaboration and collaborative work (Liu et al., 2021). Several studies have shown that there is a significant positive correlation between the degree of scientific research collaboration and the academic impact of research output. Different patterns of collaboration provide different innovative raw materials, including background knowledge, research methods, data sources, and financial support, which are conducive to promoting the generation of major innovations through resource integration (Iwasaki, 2019). Secondly, the development of the collaboration pattern is conducive to the refinement of the division of labor in scientific research collaboration, shortening the time of scientific innovation, increasing the output of scientific results, and promoting the overall process of scientific discovery (Haeussler,2020). An in-depth understanding of the relationship between scientific research collaboration patterns and scientific innovation is of great significance in guiding the efficient formation and continuous optimization of scientific and technological teams, and in promoting major breakthroughs in the fundamental research in the field of library and information science in China to achieve original results.

Therefore, this study calculates the novelty of representative papers in the field of LIS in China based on the theory of combinatorial innovation and conducts topic modeling analysis of each paper. This paper aims to address the following three research questions:

***RQ1***: What are the specific research topics of academic papers in the field of LIS in China from 2000 to 2022?

***RQ2***: How is the distribution characteristic of novelty in research papers within different topics in the field of Chinese library and information science?

***RQ3***: What is the difference in the pattern of collaboration between research papers with different topics and different degrees of novelty in the field of Chinese library and information science?

# 2. Literature review

This section mainly introduces relevant research on topic modeling analysis and the measurement of academic paper novelty in the field of Chinese library and information science.

***2.1 Novelty measurement method for academic papers***

Innovation drives sustainable development and constitutes the core essence of scientific inquiry (Park et al., 2023). The innovative capacity of a paper symbolizes the level of academic influence and the direction of research development, with novelty being a crucial aspect of evaluating paper innovation (Luo et al., 2022). Methods predominantly employed for assessing the novelty of papers include qualitative evaluation and quantitative measurement. Qualitatively, peer review, as a significant method, plays a pivotal role in assessing the novelty of papers. Quantitatively, the academic community gauges paper novelty from two perspectives: first, novelty measurement based on content analysis, and second, novelty measurement based on citation analysis (Huang et al., 2022).

**(1) Novelty measurement Based on content analysis**

Meng et al. (2023) utilized a method employing word lists, such as the semantic similarity calculation method based on WordNet, to normalize and semantically expand the keywords provided by the authors, thereby computing textual similarity to determine novelty. Zhang et al. (2021) proposed a novelty determination method based on word overlap, assessing the novelty of literature by examining the degree of textual overlap, where lower overlap indicates greater novelty. Yang et al. (2022) commenced the research topic by utilizing the Key graph algorithm to extract keywords. Subsequently, they calculated the similarity between the paper's topic and the research frontier, ultimately integrating external indicators such as journal impact factor and Altimetric to evaluate the innovation of papers comprehensively. Shibayama et al. (2021) proposed a new method to measure the novelty of scientific articles based on citation data and text data. The proposed method considers an article to be novel if it cites a combination of semantically distant references. To do this, they first assigned each cited reference a word embedding—a vector representation of each word—based on the textual information contained in the reference. Based on these vectors, the distance between each pair of references is calculated. This approach leverages limited textual information (titles of references) and a publicly shared word embedding library, minimizing data access requirements and computational costs.

**(2) Novelty measurement Based on citation analysis**

Uzzi et al. (2013) posit that scientific innovation arises from the original combination of inspiring insights. From the perspective of knowledge recombination, innovativeness can be defined as the novel reintegration of existing knowledge in unprecedented ways, a viewpoint endorsed by scholars across various disciplines. They employed Monte Carlo models to generate random citation networks and studied the novelty and conventionality of citation combinations within these networks. They used co-citation journal relationships to determine whether any pair of cited references in the citation network exhibited novelty. Wang et al. (2017) enhanced the method proposed by Uzzi et al. for measuring combinational novelty by incorporating journals as constituents of knowledge. They assessed the novelty of papers by examining whether the journal pairs of cited references had never appeared in previous publications. Lee et al. (2015) similarly refined the Uzzi method. Initially, they calculated the commonness of journal pairs across the entire paper database and adopted multi-year time windows instead of single years. Subsequently, they calculated the novelty of papers based on the references of sample papers, significantly reducing computational costs. Verhoeven et al. (2016) combined insights from existing economics and management literature to describe inventions ex-ante from two dimensions of technological novelty: recombination novelty and knowledge origin novelty. They proposed operationalizations for each dimension using patent classification and citation information and evaluated the effectiveness of the proposed measures.

In summary, content-based metrics excel in accurately revealing the primary content of articles, but may not comprehensively reflect the novelty of the substantive content of papers as the variation in terminologies may often only represent superficial innovations. On the other hand, citation-based novelty measurement methods leverage extensive citation data to enhance the scientific rigor of measurement results, a feat not achievable through content analysis alone (Bornmann et al. 2019).

### *2.2 Topic modeling and novelty measurement in LIS*

Topic modeling (An et al., 2015)and novelty measurement have become important topics in the field of library and information science research because they are directly related to the exploration of academic research frontiers (Anand and Gaurav, 2024), the direction of disciplinary development, and the quality of academic achievements (Järvelin and Vakkari, 2022). Through modeling the topics of papers, researchers can systematically understand the research dynamics and internal connections of various topic areas, providing guidance and inspiration for scholars (Zhao and Zhu, 2023; Chong and Chen, 2021). At the same time, measuring the novelty of papers helps evaluate the level of innovation and contribution to academic research, providing an important basis for academic evaluation.

Luan et al. (2023) identified disciplinary research hotspots through the knowledge production and dissemination aspects of research topics. They conducted empirical analysis using research literature and citation references in the library and information science field from 2017 to 2021 as an example. They found that constructing a thematic momentum model based on indicators of knowledge production and dissemination related to research topics effectively identified current-stage disciplinary research hotspots. Cao et al. (2021) selected papers from 11 CSSCI journals in the field of information science over the past 20 years as samples. They combined LDA topic modeling with SVM classification algorithms to identify latent topics in abstracts and evaluate paper innovativeness from the perspective of topic terms. Statistical methods were employed to validate the accuracy of the evaluation results. The findings demonstrated the ability to effectively identify papers with innovative value in different periods of information science. This approach could provide reference for scholars in selecting research topics, evaluating the innovativeness of paper topics, and reviewing papers for journals. Han et al. (2020) utilized the Latent Dirichlet Allocation model as the basis for investigating the evolutionary trajectory of research topics within the field of Library and Information Science spanning the years 1996 to 2019. This study established a dynamic journal roster, comprising preeminent journals for each temporal segment. Subsequently, the titles, abstracts, and keywords of articles from these journals were aggregated as textual corpora for LDA analysis, aiming to discern foundational topics. For each epoch, the top 10 topics along with their respective keyword distributions were identified. The research findings delineate a discernible trend showcasing the gradual waning of library science's hegemonic status within the broader LIS spectrum. Figuerola et al. (2017) provided an overview of bibliometric research in the field of LIS, aiming to examine topic boundaries, major areas, and research trends from a multidisciplinary perspective. They employed the Latent Dirichlet Allocation topic model to analyze primary topics and categories within the literature corpus. Quantitative results revealed the presence of 19 significant topics, which could be categorized into four major domains: processes, information technology, specific domains within libraries, and information applications. Yang et al. (2024) employed the emerging BERTopic model to extract and identify topics from literature related to Information Resource Management (IRM) in the Web of Science database spanning the years 2013 to 2022. They combined relevant topic terms and topic distances to delineate research directions. Additionally, they utilized dynamic topic modeling to reveal the evolutionary process of the IRM field abroad.

Liang et al. (2022) employed the field of LIS in China as a case study to measure the novelty and conventionality of papers, and explored their impact on the academic influence of papers, thereby revealing the patterns of academic innovation. They measured the novelty and conventionality of 70,207 research papers in the field of library and information science included in the Chinese Social Sciences Citation Index from 2000 to 2019, and analyzed the effects of paper novelty and conventionality on the academic impact of papers. Wang et al. (2023) utilized data from the Web of Science database comprising research papers in the Library and Information Science field. They employed the BERTopic model to extract topics from the dataset and identified interdisciplinary topics based on disciplinary diversity and cohesion. Furthermore, they analyzed the evolution of interdisciplinary topics by examining the intensity and content during the evolution process.

Overall, the current research on topics and novelty in the field of Library and Information Science tends to focus more on thematic analysis, particularly emphasizing methods for thematic modeling (Si, 2023). Research on novelty is relatively scarce and predominantly relies on content-based novelty measurements that may not comprehensively reflect the essence of papers. Therefore, this paper aims to first conduct a thematic analysis of research papers in the Chinese LIS domain from the perspective of thematic analysis. Subsequently, employing citation analysis methods to calculate the novelty of research papers under different topics, the study seeks to understand novel research outcomes across various thematic areas (Cao, 2023). This approach is expected to stimulate scholars' creative thinking, provide guidance in navigating the development trajectory of the discipline, and promote sustained innovation and progress in the LIS field. Finally, the factors influencing the novelty of a research paper are a complex issue, which may be a combination of several factors such as academic age of the authors, number of authors, gender of the authors, institutional collaboration, and research topic. In this paper, after analyzing the differences in the novelty of papers with different topics, we shift the factor analysis of the differences in novelty to the topic level, and analyze the differences in the publication characteristics of papers with different topics-collaboration situation, so as to explain the reasons for the differences in the novelty of papers with different topics.

## 3. Methodology

This section first introduces the data sources of this study, and then separately introduces the topic model and the method used to calculate the novelty scores of academic papers in this study. Figure 1 shows the framework of this study.

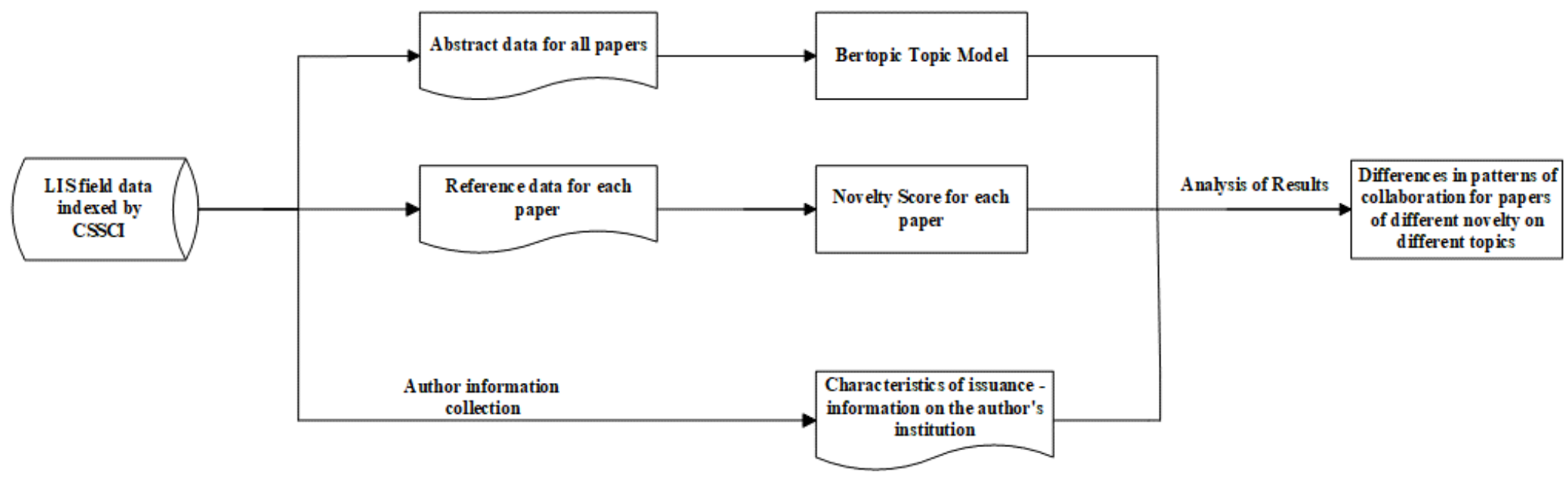

**Figure 1. Framework of this study**

### *3.1 Data collection*

The data utilized in this study originates primarily from the "Chinese Social Sciences Citation Index" (CSSCI, http://cssci.nju.edu.cn/). Managed by the Library of the Chinese Academy of Social Sciences, CSSCI is a crucial citation indexing tool that houses a vast collection of academic journal articles within the domain of library and information science. It offers comprehensive citation details, including author names, article titles, journal names, publication dates, citing authors, citing article titles, citing journals, and more. As an authoritative citation index in the library and information science field, CSSCI is highly regarded for its credibility and representativeness.

For this study, research articles were retrieved from the latest version of CSSCI's catalog of Chinese library and information science journals. The timeframe spanned from the year the journal was included in the CSSCI index to 2022, resulting in a total of 74, 988 retrieved articles. Additionally, abstract data corresponding to these articles were obtained from the China National Knowledge Infrastructure (CNKI, https://www.cnki.net/) during the subsequent phase of topic model construction. Subsequently, after excluding articles lacking publication dates, references, abstracts, and those classified as cover articles, the effective dataset for this research comprised 44, 795 articles, as outlined in Table 1.

**Table 1. Journal distribution of data sources**

| Journal | # paper | Journal | # paper |
|---|---|---|---|
| *Journal of Academic Libraries （大学图书馆学报）* | 1 967 | *Library Development(图书馆建设)* | 3 233 |
| *Archives Science Bulletin(档案学通讯)* | 2 349 | *Library Tribune(图书馆论坛)* | 4 972 |
| *Archives Science Study(档案学研究)* | 2 122 | *Research on Library Science(图书馆学研究)* | 2 866 |
| *Journal of the National Library of China(国家图书馆学刊)* | 1 047 | *Library Journal(图书馆杂志)* | 4 050 |
| *Information Science(情报科学)* | 6 744 | *Library and Information Service(图书情报工作)* | 1 809 |
| *Information Studies: Theory & Application(情报理论与实践)* | 5 401 | *Documentation, Information & Knowledge(图书情报知识)* | 2 394 |
| *Journal of the China Society for Scientific and Technical Information(情报学报)* | 1 401 | *Library & Information(图书与情报)* | 2 532 |
| *Journal of Intelligence(情报杂志)* | 2 323 | *Journal of Modern Information(现代情报)* | 1 374 |
| *Information and Documentation Services(情报资料工作)* | 1 897 | *Journal of Information Resources Management(信息资源管理学报)* | 195 |
| *Data Analysis and Knowledge Discovery(数据分析与知识发现)* | 663 | *Journal of Library Science in China(中国图书馆学报)* | 1 065 |

### ***3.2 Topic analysis***

This study employs a deep learning-based approach to uncover and analyze the evolving topics within the field of LIS in China, focusing on abstracts of research papers. BERTopic stands out as a topic modeling technique leveraging BERT word embeddings, empowered by Transformer and c-TF-IDF(continuous-Term Frequency-Inverse Document Frequency) to create dense clusters that not only simplify topic interpretation but also retain crucial terms within topic descriptions(Abuzayed and Al-Khalifa, 2021). c-TF-IDF is a text feature extraction method, an extension of the TF-IDF method. The characteristic of c-TF-IDF lies in its consideration of the positional information of words, meaning that words in different positions have different importance for the text. The calculation method of c-TF-IDF is similar to TF-IDF, but it takes into account the positional information of words when calculating term frequency (TF) and inverse document frequency (IDF), thereby better capturing the importance of words. Its implementation spans three key stages (Grootendorst, 2022):

Embedding documents: Utilizing BERT or alternative embedding techniques to extract document embedding vectors.

Clustering documents: Employing the UMAP (Uniform Manifold Approximation and Projection) to reduce embedding dimensionality, followed by HDBSCAN (Hierarchical Density-Based Spatial Clustering of Applications with Noise) application to further condense embeddings and form clusters of semantically similar documents. UMAP is a dimensionality reduction algorithm commonly used to visualize high-dimensional data. The core principle of UMAP is based on the preservation of local continuity structure, which maps high-dimensional data to a low-dimensional space through the construction of neighborhood graphs and optimization processes, so that the local relationships of the data are still preserved after dimensionality reduction.HDBSCAN is a density-based clustering algorithm that can identify clusters of arbitrary shapes and sizes in data sets. It works by constructing a minimum spanning tree based on density reachable graphs and then applying density-based clustering techniques, subsets with high densities are found as clusters and data points that cannot be categorized are marked as noise points.

Representing topics: Calculating vocabulary importance within clusters using the c-TF-IDF algorithm, an enhanced version of TF-IDF, and subsequently extracting candidate words most relevant to topics based on maximum marginal relevance.

In the realm of topic modeling algorithms, the BERTopic model exhibits distinct advantages over traditional approaches such as Latent Dirichlet Allocation (LDA) (Blei et al., 2003) and Topic2Vec (Niu et al.,2015) models. Leveraging the Transformer architecture based on the BERT model, BERTopic can capture semantic relations and contextual information among words more accurately, thus generating richer and more semantically relevant text representations with strong sequence modeling capabilities. Given the substantial volume of data in this study and the lengthy and complex nature of academic abstracts, BERTopic proves to be effective in extracting topics from the paper within the field of LIS in China, identifying key research areas, and discerning the evolution and trends in these topics, thereby facilitating analysis of their novelty.

The detailed parameter settings for conducting topic modeling using BERTopic in this study are as follows(Chagnon et al., 2024):

(1) Embedding Documents:

The document embeddings are generated using the embedding model "paraphrase-multilingual-MiniLM-L12-v2."

(2) UMAP Initialization

UMAP is initialized with the following parameters based on multiple rounds of experimentation and relevant literature: the projected dimensionality (n-components) is set to 50; the distance between data points is measured using cosine similarity; and the minimum distance between points is adjusted from the default value of 0.1 to 0.001 to achieve denser embeddings.

(3) HDBSCAN Initialization

HDBSCAN is initialized with the following parameters: the minimum cluster size is set to 50, indicating that a cluster must contain at least 50 samples; the minimum number of samples is set to 25, indicating that a data sample must have at least 25 samples in its neighborhood to be considered a core point; otherwise, it will be treated as noise. Here, core points refer to data samples that are considered high-density regions in density clustering.

(4) Topic Extraction

In this study, we used uni-gram, which means that only single words are considered as features. This means that during the topic extraction process, Bertopic analyzes each individual word as a feature without considering combinations between words. This approach simplifies the feature space while retaining enough semantic information and usually yields good topic extraction results. Additionally, considering prior research on topic modeling in the domain of LIS (Zhang C et al.,2023), the number of topics is set to 20, and Bertopic will perform clustering based on the input text data and parameters to obtain the classification results.

### *3.3 Novelty measurement*

In this study, we utilized method of Lee et al.(2015), which is grounded in the theory of combinatorial innovation, to quantify the novelty of research papers. This method, derived from a modified version of Uzzi et al.(2013) measurement approach, views a paper's novelty through the lens of the rarity of pairs of references. Its premise lies in the observation that certain knowledge domains exhibit rarer or more innovative pairs of combinations compared to others. Given that each paper draws from diverse prior knowledge sources, it generates a knowledge combination distribution chart for each paper. Subsequently, the distribution's characteristics can be described, such as relative scarcity or commonality at various cut-off points (e.g., median or 10th percentile).

Compared to Uzzi's Monte Carlo simulations for constructing multiple random citation networks to compute the expected frequency of journal pairs, Lee(2015) employs probabilistic statistical methods to calculate the expected frequency of each journal pair. This reduces computational complexity, rendering it more suitable for this study given its computational demands.

The novelty measurement process comprises two primary steps (Lee,2015):

(1) calculating the commonality of co-cited journal pairs across the database; and (2) calculating the novelty of papers based on the references of our papers.

The first step performs the following procedure: (1) retrieve all papers in the field of LIS in the cssci index; (2) retrieve all references for each paper; (3) list all paired reference combinations for each paper; (4) record the two paired journals for each pair of references; (5) record the year of publication, t, for each paper; and (6) construct a range of paired journals for each year by pooling together all paired journals published in the same year. range of paired journals for each year. We refer to the journal pairs in year t as Ut. We then define the degree of commonality for each pair (journals i and j) in year t as:

$$Commonness_{ijt} = \frac{N_{ijt}}{\frac{N_{it}}{N_t} \cdot \frac{N_{jt}}{N_t} \cdot N_t} \tag{1}$$

Where $N_{ijt}$ denotes the number of journal pairs of $i$ and $j$ within $U_t$, $N_{it}$ signifies the count of journal pairs in $U_t$ containing journal $I$, and $N_t$ stands for the total count of all journal pairs in the entire dataset.

The second step calculates the novelty at the paper level. For papers published in year t, we record their commonality per pair of cited journals. This produces a series of commonality values, and we subsequently sort these numbers and record the 10th percentile as an indicator of paper-level commonality. In addition, taking the 10th percentile rather than the minimum value reduces noise and increases the reliability of the measure.

# 4. Results

This study employs Chinese scholarly journals in the field of graphic information indexed by CSSCI as the primary data source. Initially, the BERTopic model is applied to conduct topic modeling on the research papers. Subsequently, novelty scores are computed for all papers to derive the novelty distribution across different topics. Further analysis incorporates a temporal dimension to demonstrate the evolution of novelty scores for different topics over time. Finally, differences in author collaboration patterns between papers of varying novelty were investigated at the topic level.

***4.1 Results of topic analysis of scholarly papers in the field of LIS in China***

We answer RQ1 in this section. This paper conducts topic classification modeling on abstracts of Chinese literature in the field of library and information science. Based on the input hyperparameters and data features, a total of 15 topic categories were obtained by training the BERTopic topic model and merging similar topics automatically. Additionally, 26, 891 papers were considered noise (composed of outliers and topics with fewer than 50 papers) and labeled as -1. The hierarchical clustering of merged topic categories is illustrated in Figure 2, demonstrating clear hierarchical relationships between various topics. For instance, topic 0 -library-reading-service and topic 4 -China-library-library science exhibits the most closely related hierarchical structure. Figure 3 presents the top five feature words for each topic in the form of a bar chart, and topic labels are determined based on relevant topic words and their c-TF-IDF scores. According to the TF-IDF score ranking, select the highest score of a number of words (usually 5-10) as the topic of the representative words, these words are the topic of the keywords, can reflect the core content of the topic. The research topics in the Chinese literature field of library and information science include but are not limited to library-reading-service, data-information-user, archives-documents-archive, journal-research-evaluation, library-China-library science, intelligence-competition-security, culture-public-service, literature-blackboard-document, patent-technology-method, text-semantics-classification, search-web2-blog, virtual-community-knowledge, computing-service-security, paper-archive-handwriting, and OA-journal-open access, among others.

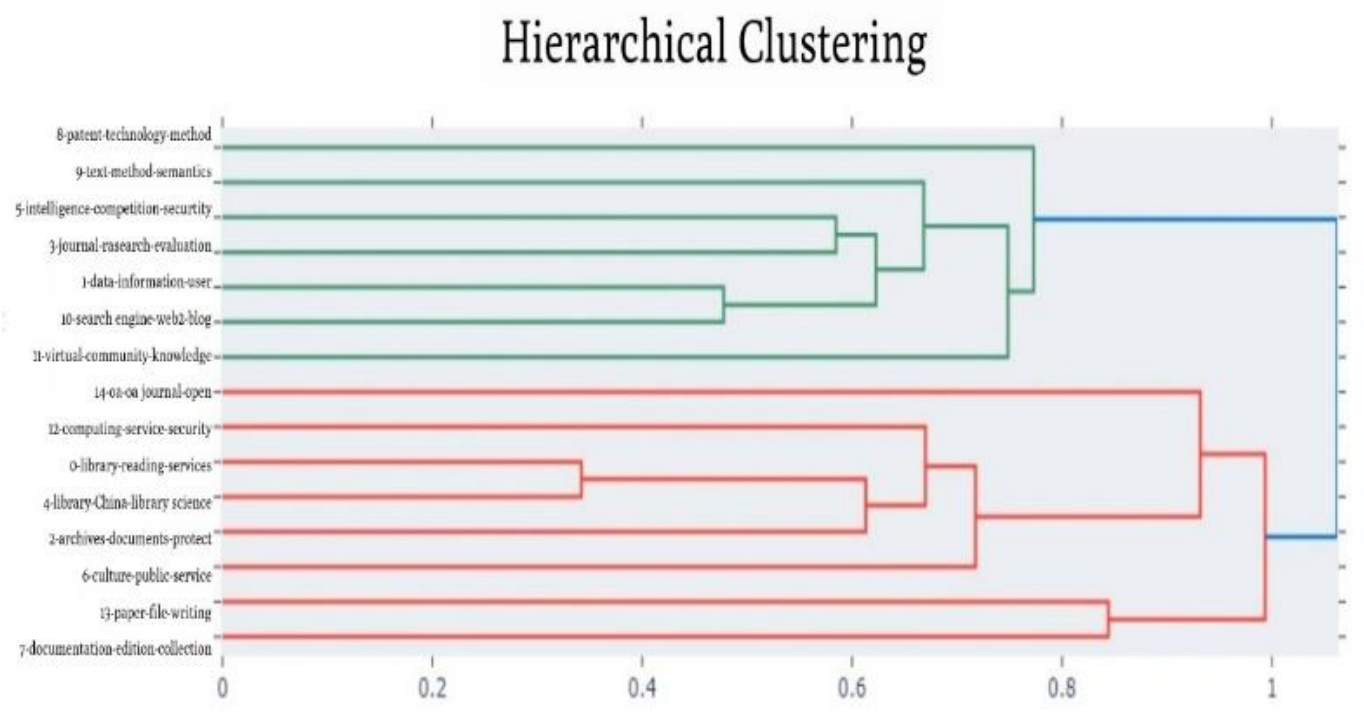

**Figure 2. Latent hierarchy of topic distribution**

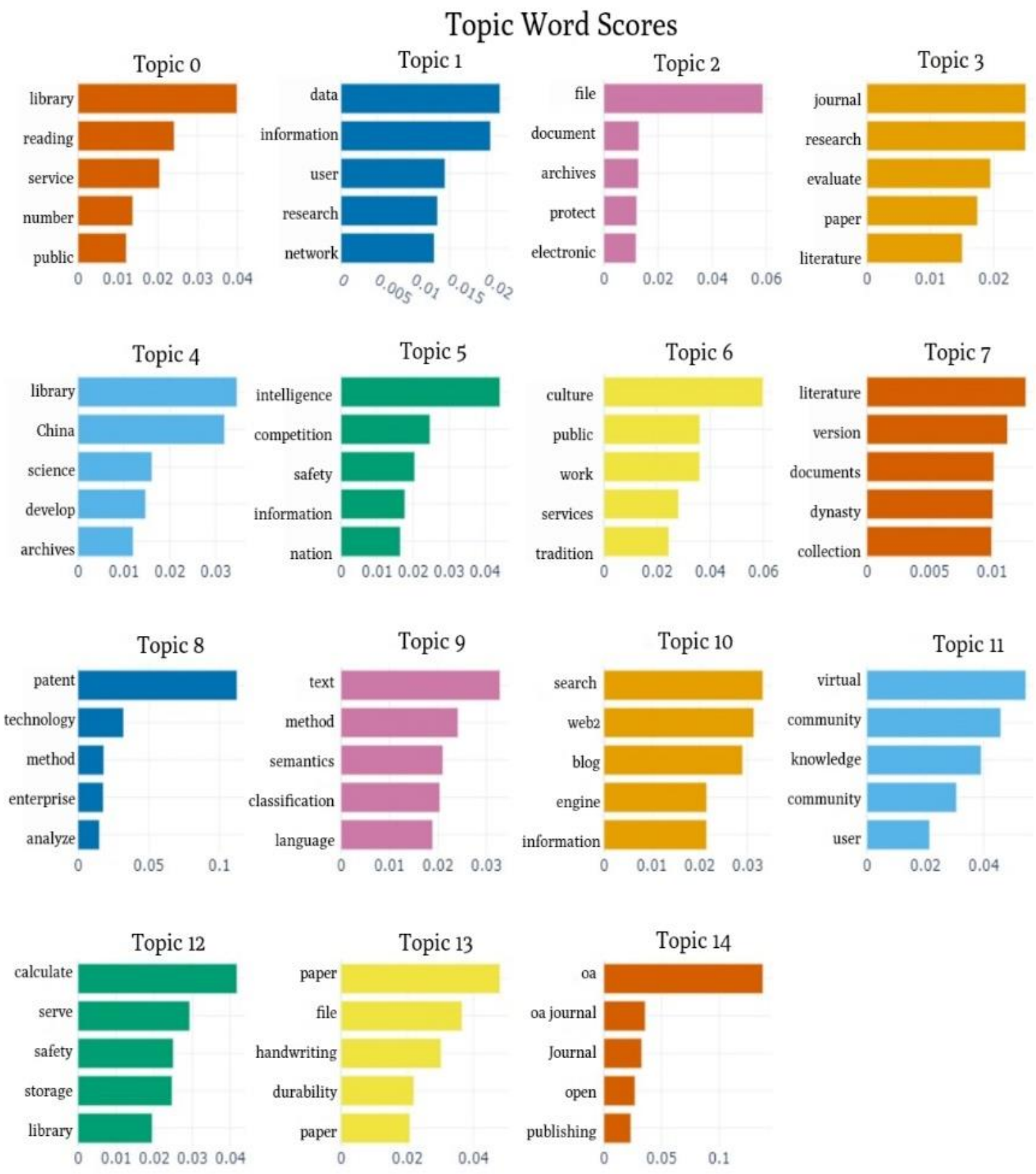

**Figure 3. Topic feature words for each topic**

To provide a more intuitive representation of the document distribution characteristics within each topic, we have generated a document-topic distribution graph, as shown in Figure 4. Each cluster of the same color represents documents belonging to the same topic. It can be observed that in recent years, research papers in the field of LIS in China primarily focus on topics such as topic0-library-reading-service, topic1-data-information-user, topic3-journal-research-evaluation, and topic5-intelligence-competition-security.

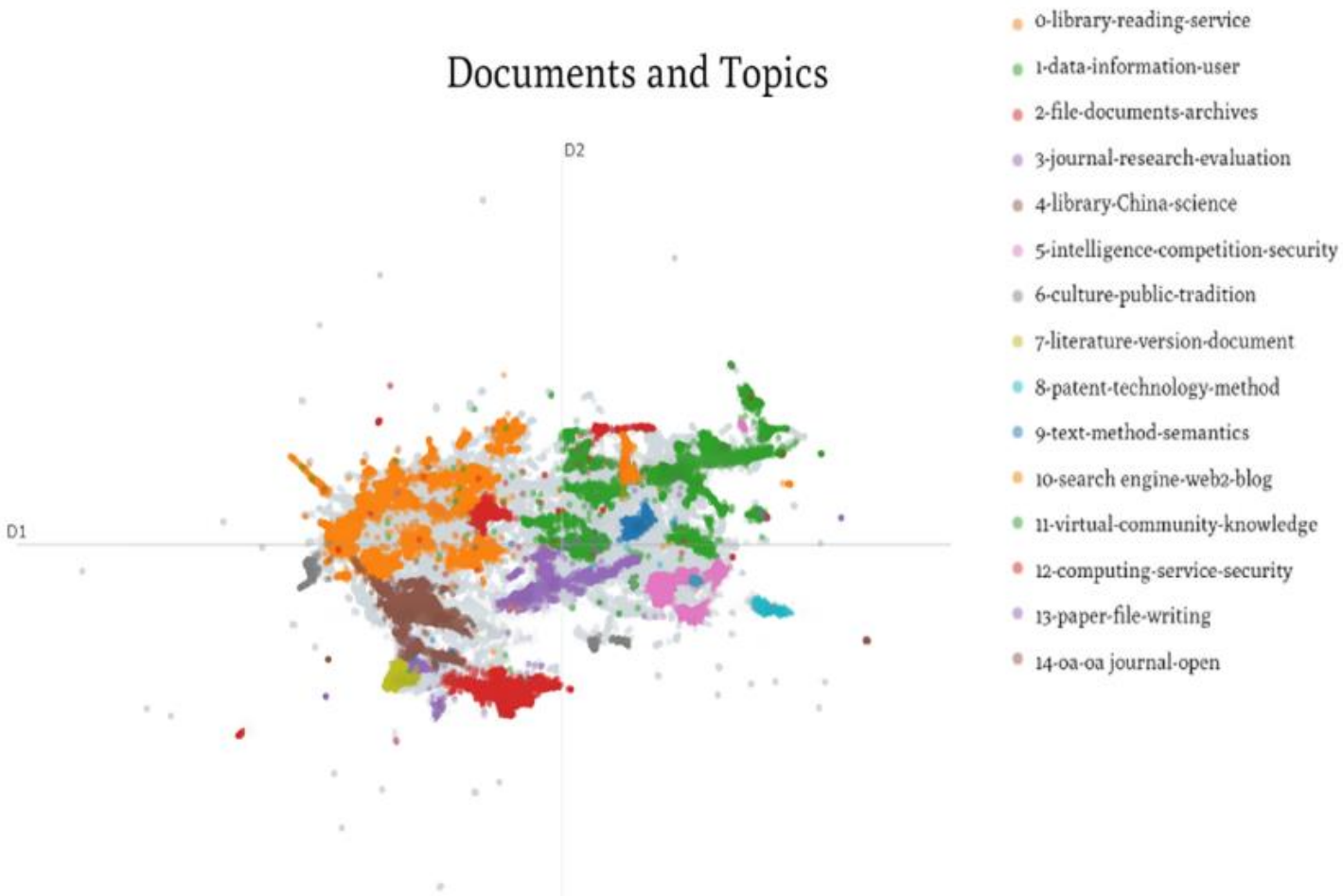

**Figure 4. Document - Topic Distribution Map**

In addition, this study computed coherence scores (Newman, D et al.,2010) for each topic to validate the effectiveness of the topic model classification. Topic coherence is commonly used to quantify the consistency and coherence of topics extracted by topic models, reflecting the semantic relatedness among the words or documents within each topic. Specifically, the formula is as follows:

$$Coherence_score = \frac{1}{|\{(\omega_i,\omega_j)\}|}\sum_{(\omega_i,\omega_j)} log\frac{P(\omega_i,\omega_j)}{P(\omega_i)P(\omega_j)} \quad (2)$$

Where $|\{(\omega_i,\omega_j)\}|$ represents the total number of word pairs,$P(\omega_i,\omega_j)$ represents the probability of words $\omega_i$ and $\omega_j$ co-occurring in the same document, and $P(\omega_i), P(\omega_j)$ represent the probabilities of words $\omega_i$and $\omega_j$ appearing independently.

Generally, higher coherence scores indicate greater semantic consistency and coherence within the extracted topics. The results are presented in Table 2. It can be observed that the coherence scores for each topic range from 0.44 to 0.83, suggesting overall good quality of these topics. Specifically, the highest coherence score is achieved by Topic13-paper-file-writing with a score of 0.833, followed by Topic11-virtual-community-knowledge with a score of 0.711, indicating a high degree of semantic relatedness among the words or documents within these topics. However, some topics exhibit relatively lower coherence scores, such as Topic2-archives-documents-archive, with a score of 0.440. This may suggest weaker semantic relatedness among the words or documents extracted within this topic, or insufficiently distinct features captured by the topic model.

Overall, the BERTopic topic model used in this study for topic modeling of the Chinese LIS field has achieved excellent results, effectively reflecting the distribution of research topics in the LIS domain.

**Table 2. Consistency scores for each topic**

| Topic | Consistency Score |
|---|---|
| 0 | 0.690 |
| 1 | 0.589 |
| 2 | 0.440 |
| 3 | 0.529 |
| 4 | 0.510 |
| 5 | 0.526 |
| 6 | 0.513 |
| 7 | 0.604 |
| 8 | 0.624 |
| 9 | 0.530 |
| 10 | 0.509 |
| 11 | 0.711 |
| 12 | 0.493 |
| 13 | 0.833 |
| 14 | 0.687 |

***4.2 The overall novelty of scholarly papers in the field of LIS in China***

This study calculates the novelty of all research papers in the field of LIS in China based on the improved novelty calculation method proposed by Lee. The distribution of novelty scores is shown in Figure 5. From the figure, it can be observed that the distribution exhibits a clear skewness, specifically a left-skewed distribution. Regarding the novelty of research papers in the field of LIS in China, the majority of scores lie between 0 and 1, followed by scores between -1 and 0. The sum of the proportions of these two items reaches 59.07%, indicating that a minority of papers exhibit high novelty while another minority exhibit low novelty, which aligns with previous research and expectations.

However, it should be noted that the novelty calculation method is based on the combination of reference journal pairs, and due to the limited number of authoritative Chinese journals in the library and information science field, the highest novelty score in the entire dataset is only around 3, while the lowest score reaches approximately -12.5. This limitation in the availability of Chinese authoritative journals may have influenced the distribution of novelty scores observed in this study.

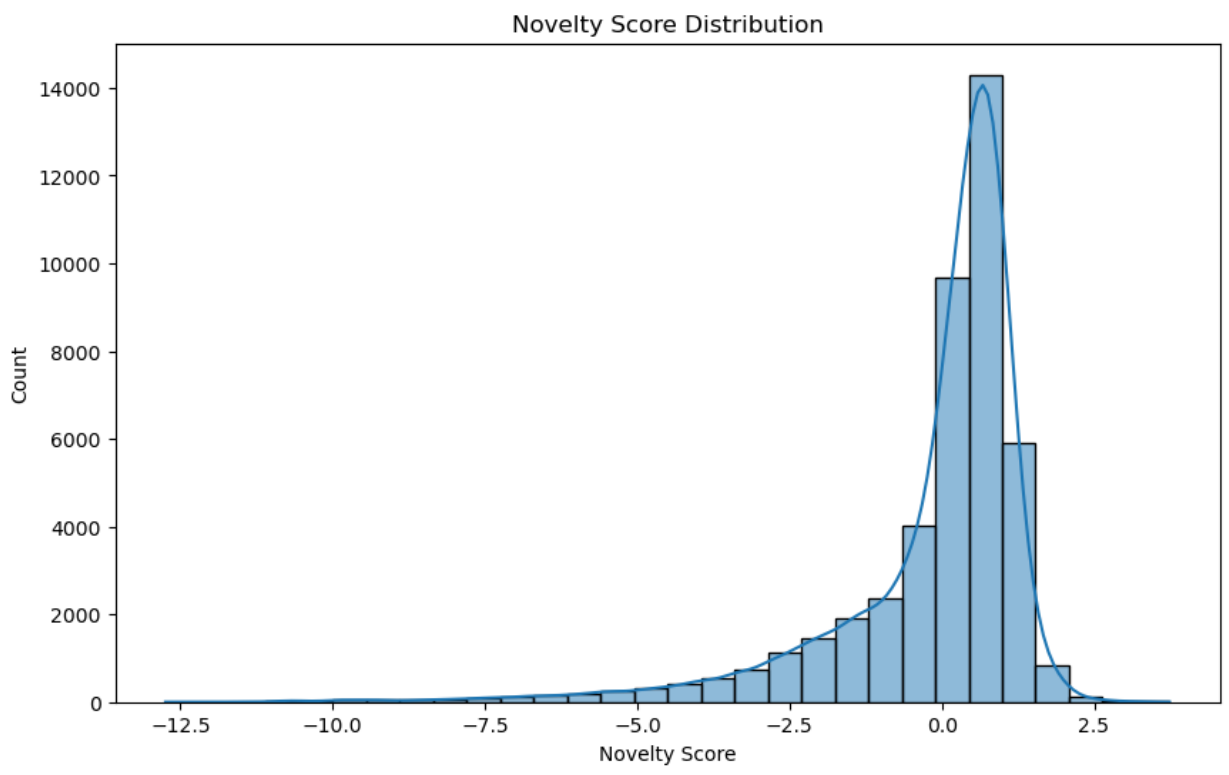

**Figure 5. Overall distribution of novelty scores**

After calculating the novelty scores for all research papers, we first analyzed the distribution of novelty scores for papers published in different journals to gain preliminary insights into the distribution of novelty among journals in the field of LIS in China. We plotted the scores using a box plot, as shown in Figure 6.

From the plot, it can be observed that Archives Science Bulletin, Archives Science Study, and Library Journal exhibit the most scattered distribution of novelty scores among the three journals analyzed. Additionally, the median novelty scores for papers published in these journals are noticeably lower than other journals, suggesting that these journals feature both highly novel and less novel papers, resulting in a lower novelty score than other journals. Journals with higher overall novelty scores include the Journal of the China Society for Scientific and Technical Information, Journal of Information Resources Management, and Journal of Modern Information. The reason for the above phenomenon may be attributed to the fact that archival science is a field that emphasizes literature, archival management, and information organization, with relatively stable research topics and methods. Additionally, since archival science focuses on historical documents and archival materials, which are inherently historical in nature, this limits the novelty of papers to some extent.

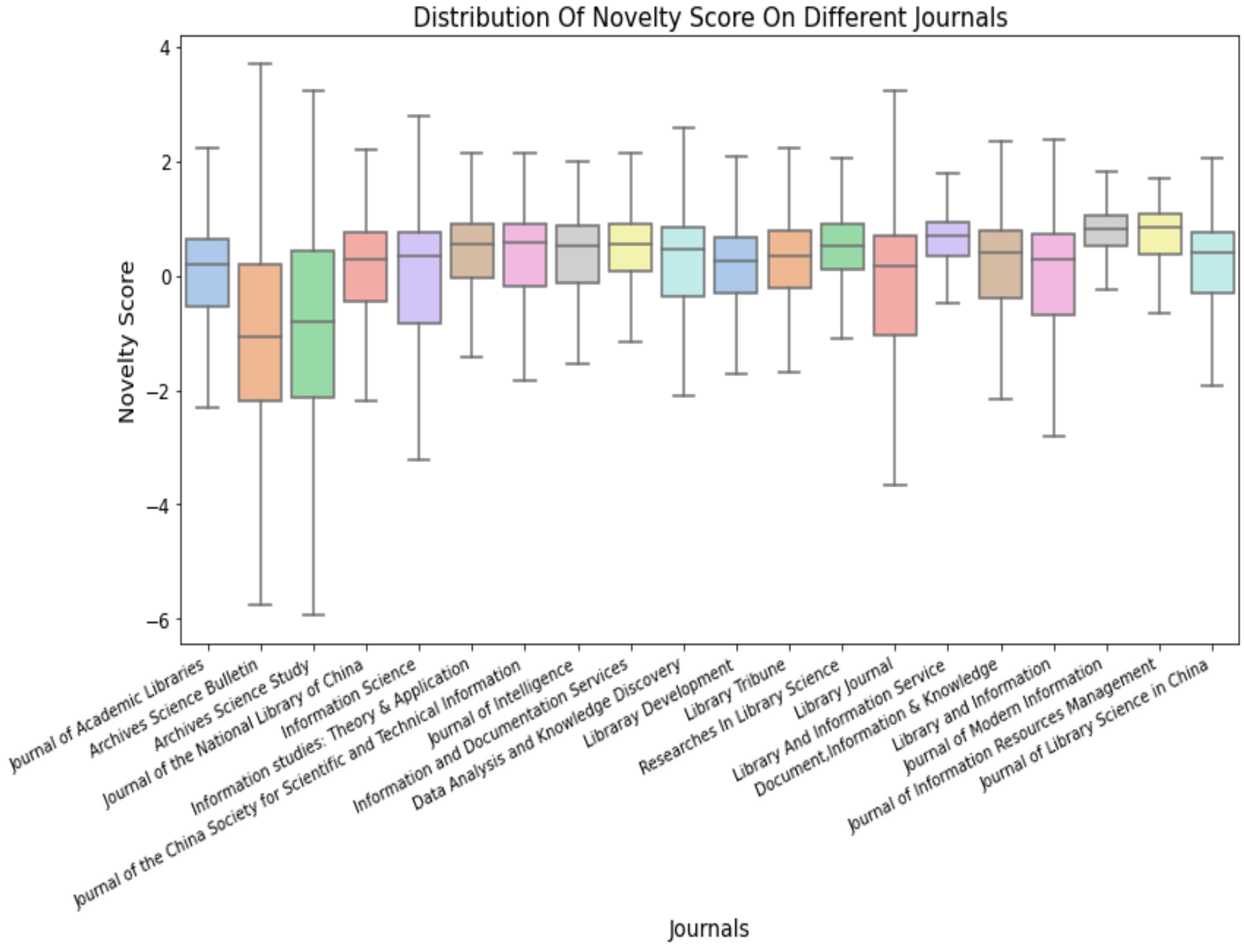

**Figure 6. Distribution of novelty scores in different journals**

### *4.3 The novelty of scholarly papers in the field of LIS in China with different topics*

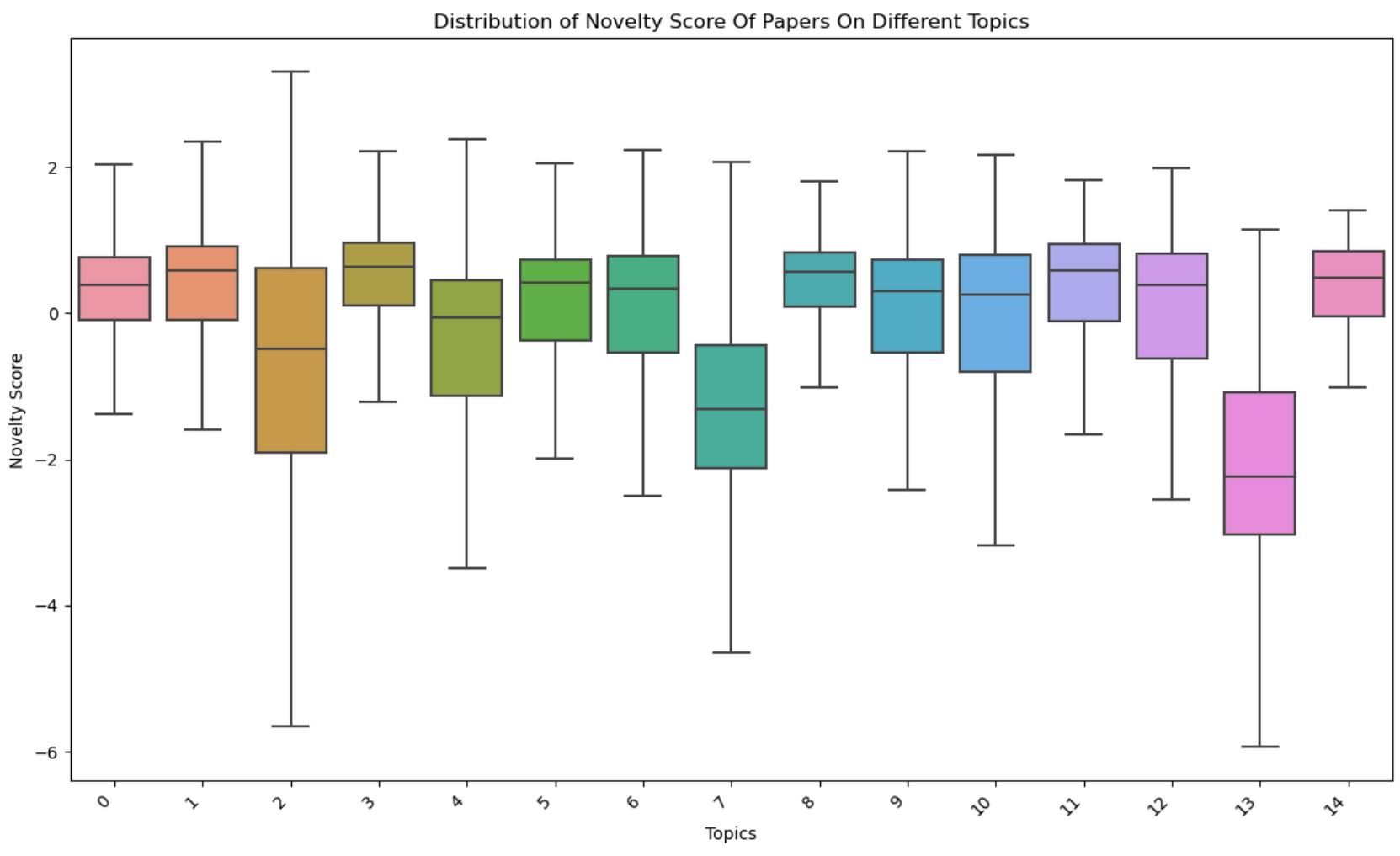

**Figure 7. Distribution of novelty scores for different topics**

We answer RQ2 in this section. We combined the previous topic model analysis results with the no velty scores to analyze the distribution of novelty scores for papers across different topics. The res ults are depicted in Figure 7. Figure 7 shows that the median novelty scores for papers in topic13-p aper-file-writing, topic7-literature-version-document, and topic2- archives-documents-archive are significantly lower than those for other topics. After computation, the average novelty scores for th ese three topics are as follows: -2.246834, -1.367731, and -0.785096, respectively. Combining thes e results with the distribution of journals and their novelty scores, it is apparent that these three top ics align with the research directions of Archives Science Bulletin, Archives Science Study, and Li brary Journal. This suggests that in the field of LIS in China, papers focusing on topics related to p aper handwriting, literature documentation, and archive documents exhibit lower novelty compare d to other topics. On the other hand, topics with higher median novelty scores include topic-3-jour nal-evaluation-research, topic-8-patent-technology-method, and topic-11-virtual-community-know ledge, with average novelty scores of 0.214776, 0.203686, and 0.106732 respectively. The reason behind this may be that these topics involve constantly evolving fields and issues, requiring contin uous exploration of new theories, methods, or application scenarios, thus often exhibiting higher n ovelty.

Subsequently, we selected the top 1000 papers with the highest novelty scores and the bottom 1000 papers with the lowest novelty scores for further empirical analysis of their topic distribution. The results are depicted in Figure 8. By comparing the topic distribution of papers with the highest and lowest novelty scores, we can observe that papers with the highest novelty scores mainly belong to topic0-library-reading-service and topic-4-library-China-library science. These two topics, compared to others, have broader coverage and more macroscopic research, making them influential in the long-term development of the field. Therefore, papers in these areas tend to have higher novelty. On the other hand, papers with the lowest novelty scores are mostly related to topic1-data-information-user and topic2-archives-documents-archive, further confirming that papers related to archive research topics often exhibit lower novelty.

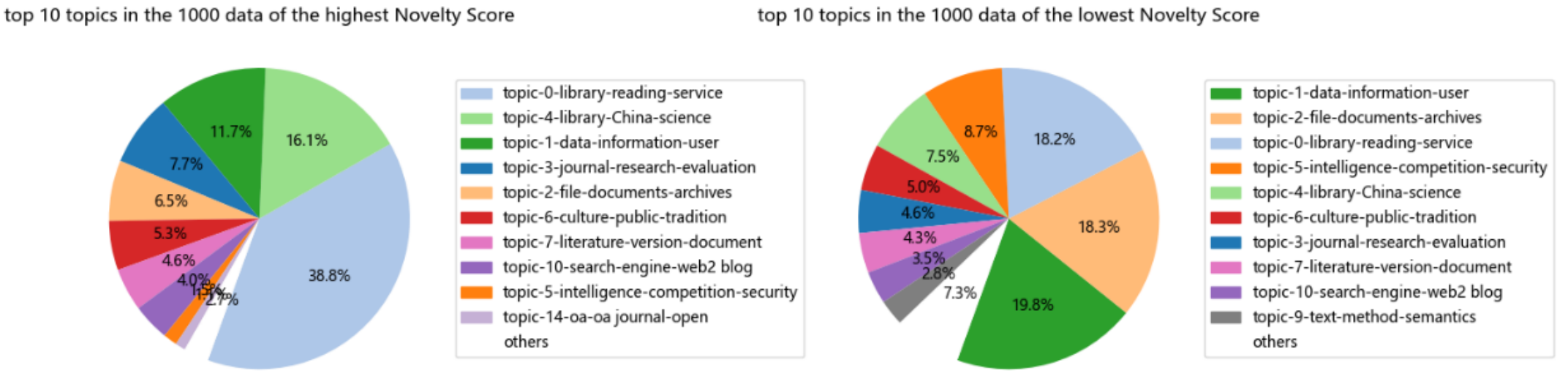

**Figure 8. Topic distribution of high novelty and low novelty papers**

Next, to assess the evolution of novelty scores across different topics over time, we divided the publication years of papers into three-year intervals and plotted the time evolution of novelty scores for papers in different topics, as shown in Figure 9. Additionally, we calculated the average novelty scores for all topics and plotted their evolution curve in the same figure.

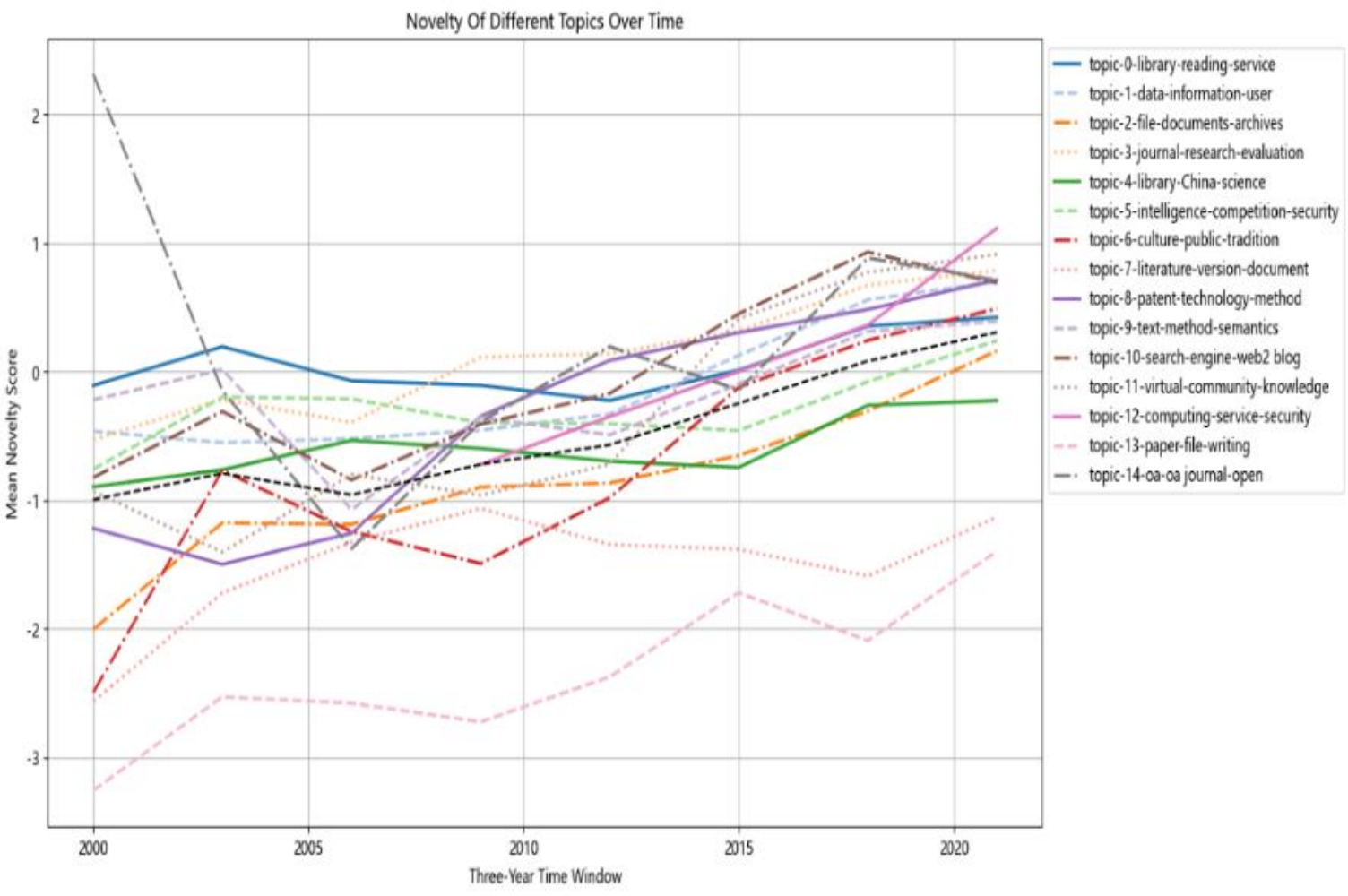

**Figure 9. The evolution of novelty under different topics**

Overall, in the field of Chinese library and information science, from 2000 to 2022, the novelty scores of research papers have remained around 0, indicating that most papers exhibit neither high nor low novelty. However, there is a slow upward trend in novelty scores over time. Specifically, topic14-OA-journal-open access had the highest novelty around the year 2000, reaching a score of over 2. Looking back on history, we can attribute this to the emergence of the open-access journal concept, which was proposed by the Budapest Open Access Initiative in the early 20th century. The subsequent decline in novelty scores from 2000 to 2006 suggests a waning research interest during that period, followed by a gradual recovery and upward trend after 2006.Topic13-paper-file-writing, on the other hand, maintained the lowest novelty scores throughout the entire period from 2000 to 2022. This could be due to the relatively traditional and stable nature of this field, where techniques and methods are relatively mature. Topics6-culture-public-service and topic8-patent-technology-method exhibited consistent growth in novelty scores over the entire time range, indicating ongoing development and evolution in these areas. This could be attributed to the interdisciplinary nature of these topics, involving fields such as cultural studies, public administration, technology, and law. The interdisciplinary integration may have facilitated knowledge exchange and collaboration between different fields. In the case of patent-technology-method, the continuous demand for innovation may have driven researchers to explore and innovate continuously, leading to the advancement of research in these areas and the trend of increasing novelty scores.

### *4.4 Distribution of Author Collaboration in Papers with Various Topics and Degrees of Novelty*

We answer RQ3 in this section. The previous results indicate that research papers in the field of LIS in China have exhibited a continuous upward trend in novelty from 2000 to 2022. This trend may be attributed to changes in various aspects, such as society, technology, or policies, leading to different shifts in research hotspots across various topics. However, existing research suggests that the factors affecting novelty of a paper are not limited to the research topic itself, but are also closely related to many other factors, making it difficult to quantitatively analyze the factors influencing novelty of a paper in terms of correlation or causality. However, there is growing evidence (Lyu, 2021) that author collaboration plays a key role in increasing the impact or novelty of scientific achievements. Collaboration not only brings about the exchange and collision of ideas, but also leads to the sharing of resources and complementary technologies, all of which may have a significant impact on the novelty of research results.

Author collaboration is a common and effective pattern in academic research. Through collaboration, researchers can combine their expertise and resources to tackle complex problems that are difficult to solve individually. In addition, inter-institutional collaboration is an important way to advance research progress, and collaboration across institutions can break down geographic and disciplinary boundaries to promote broader knowledge sharing and innovation. In order to gain a deeper understanding of the impact of these modes of collaboration on the novelty of papers in the field of intelligence, this section will focus on analyzing differences in author collaboration across topics. We will examine the number of author collaborations, the structural characteristics of the collaborations, and the collaborative relationships between different institutions by statistically analyzing the author information contained in the data of the paper's title. Meanwhile, through these analyses, we expect to reveal the distribution of the number of co-authors as well as the mode of collaboration in intelligence studies with different research topics, and to explore its importance for enhancing academic innovation after combining with the paper's novelty score.

We first start from the data of a total of 44,795 papers under all topics, and explore the effects of the number of co-authors and the pattern of collaboration on the novelty of the papers without considering the topic classification. By extracting and analyzing the names and institutional information of all the authors of the research papers, each author selects the first institution as his affiliation, and when there is only one author, the paper is an independent work, and when the number of authors of the paper is greater than one, the authors' collaboration patterns are categorized into intra-institutional and inter-institutional collaboration according to the authors' affiliations, and then, combined with the novelty scores of each paper previously calculated. The relationship between different number of co-authors and different collaboration patterns and the novelty scores of the papers were investigated, and the results are shown in Figure 10 and Figure 11.

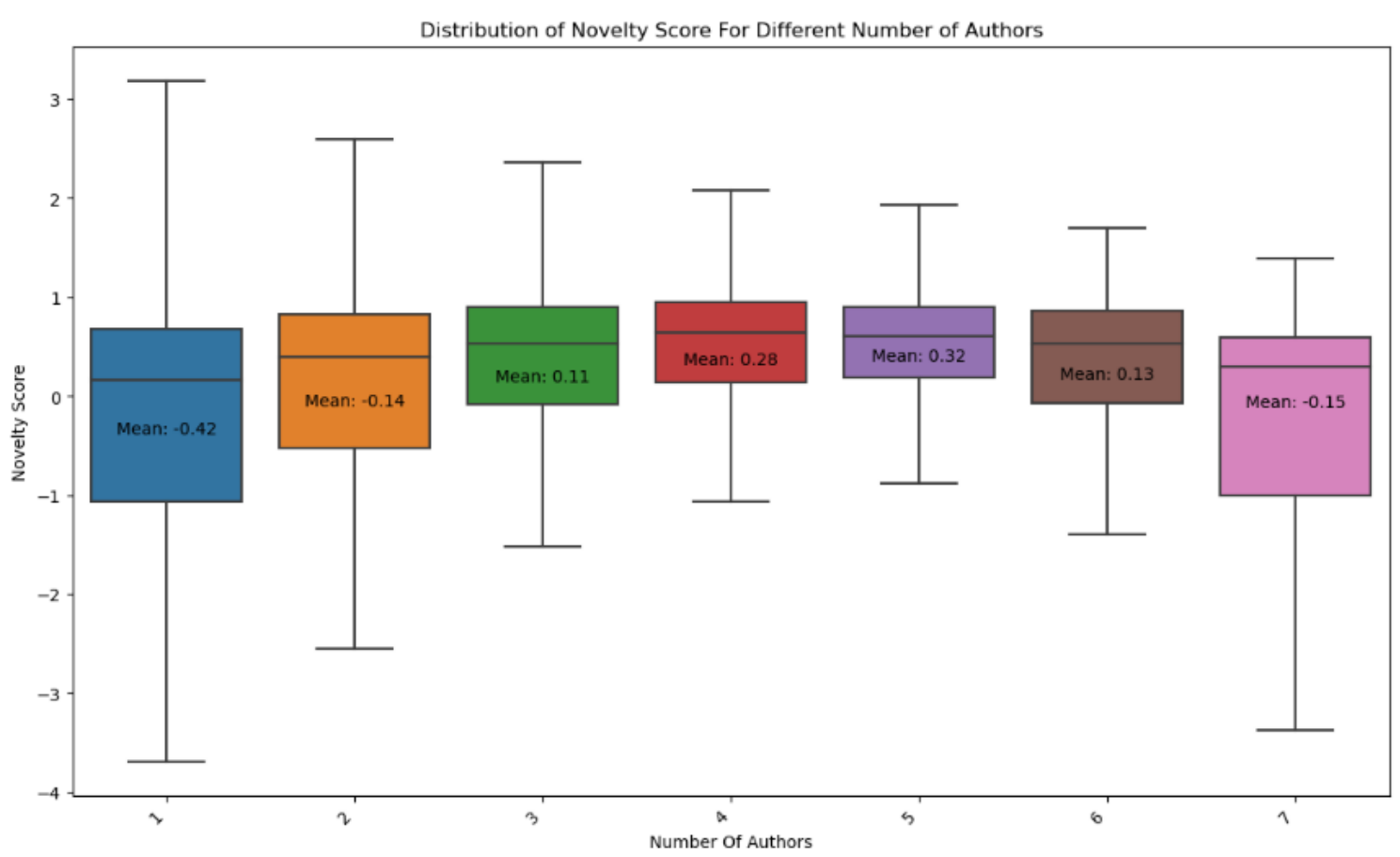

**Figure 10. Novelty distribution of papers with different numbers of co-authors**

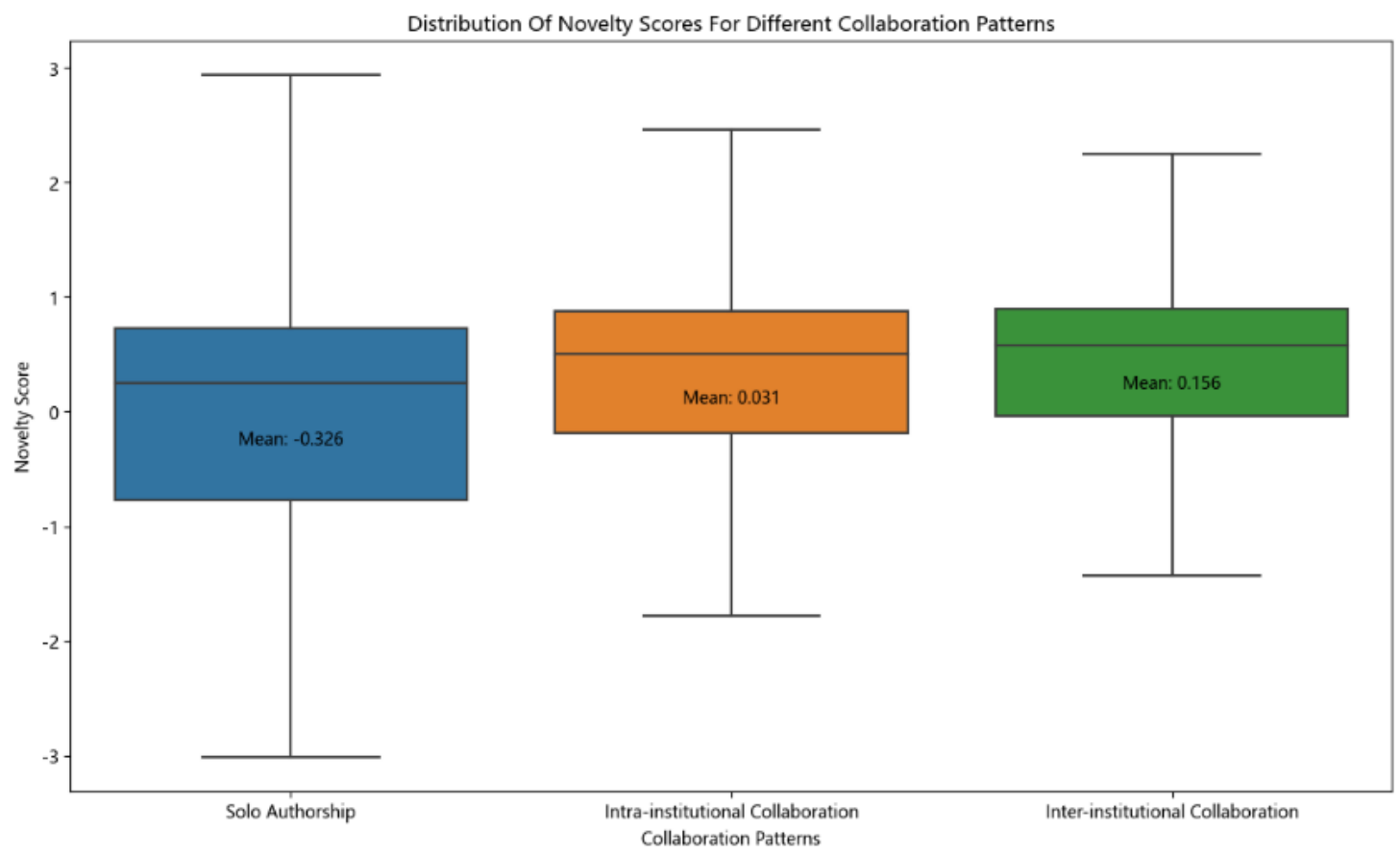

**Figure 11. Novelty distribution of papers with different collaboration patterns**

From Figure 10, we can observe that as the number of co-authors increases, both the average and median novelty scores of papers continue to rise, reaching their highest values when the number of co-authors reaches 4 and 5. Referring to previous studies, this trend suggests that papers with multiple co-authors may encompass multiple research fields or diverse professional backgrounds. Each author can contribute unique ideas and insights from different perspectives, injecting more novelty into the paper. Additionally, papers with multiple co-authors may leverage more resources and skills, including research equipment, experimental techniques, and data analysis capabilities. The richness of these resources and skills may provide greater support and assurance for the research conducted in the paper, thereby enhancing its novelty. However, as the number of authors increases, the complexity of collaboration also increases. In a large team, there may be more disagreements, communication barriers, and collaboration difficulties, making it challenging to maintain a high level of novelty.

According to Figure 11, we can see that the three patterns of solo authorship, intra-institutional collaboration, and inter-institutional co-authorship exhibit significant differences in terms of novelty. First, solo authors are often limited by their own knowledge and resources in the research process. This limitation may lead to a more homogeneous research perspective, making it difficult to break through the existing academic framework, which in turn affects the novelty of the paper. Lacking the exchange of ideas and technical complementarity with others, solo researchers may have more limited inspiration for innovation, resulting in relatively low performance of research results in terms of innovativeness. Second, intra-institutional collaborations can partially overcome some of the limitations of solo authorship by combining the expertise and skills of different researchers, thereby increasing the breadth and depth of research. However, participants in intra-institutional collaborations usually come from the same institution and may have similar academic backgrounds and research orientations. This homogenization may limit the diversity of the research, and although it can enhance a certain degree of novelty, it is still not as good as inter-institutional co-authorship. Finally, inter-institutional co-authorship exhibits the highest level of novelty, which can be attributed to having a broader range of research content and diverse research perspectives. Inter-institutional collaborations break down geographical and disciplinary boundaries, and the participants come from different academic and cultural backgrounds, which can bring in more diverse perspectives and innovative ideas. In addition, inter-institutional collaborations are often able to mobilize richer resources and technical support, thus promoting innovation in research in terms of methodology and content.

After analyzing the effects of the number of authors and the pattern of co-authorship on the novelty of papers, returning to our study, are there significant differences in this aspect between papers in different research topics under the field of Chinese library and information science, and thus is it possible that this may result in different performances in terms of their novelty scores? Is it possible that papers in topics with lower novelty scores have a greater proportion of sole authors and papers in topics with higher novelty scores have more co-authors?

To this end, we examined the differences in the collaboration patterns of research papers under different topics, as shown in Figure 12. Here, we first focus on the distributional characteristics of the collaboration patterns of papers under topics 2, 4, 7, and 13, which have lower novelty scores, and topics 1, 3, 8, and 11, which have higher novelty scores. We can see that, except for topic 13, topics 2, 4, and 7 all have a significantly higher percentage of independent authors than the average of 52.8%, amounting to 66.7%, 73.7%, and 85.6%, respectively, while topic 13 also has a slightly lower-than-average percentage of 50.3%. Topics 1, 3, 8, and 11 all have a higher proportion of collaboration, with the proportion of intra-institutional collaboration reaching 52.6%, 43.6%, 58.0%, and 59.5% respectively, significantly higher than the average of 39.2%, and the proportion of inter-institutional collaboration reaching 13.1%, 9.1%, 17.7%, and 11.1% respectively, which are also higher than the average of 8%. This suggests that although it cannot be directly causally shown that a higher proportion of solo authors leads to lower novelty scores for papers under these topics, and a higher proportion of collaborations leads to higher novelty scores, it intuitively demonstrates that, in the field of intelligence in China, papers under low-novelty topics (e.g., Topics 2, 4, 7, and 13) tend to have a larger proportion of solo authors, and papers under high-novelty topics (e.g., Topics 1, 3, 8 and 11) tend to have more collaborations.

Based on the above analysis, we can see that the topics related to archival studies, such as topics 2, 4, 7, and 13, which have a lower collaborative nature, may be due to the fact that, on the one hand, these topics are more specialized or the research is more traditional, which leads to the need for in-depth research and expression by a single author; on the other hand, it may mean that the research directions or issues of these topics have already had more research bases and results. In this case, it may be easier for a single author to conduct research and publish results within an already existing framework without the need for large-scale collaboration. Finally, and reflecting the academic culture and practices of the field, single authorship may be more common or preferred. At the same time, however, it is often the case that several of these reasons lead to papers in the field having lower novelty scores. Topics with high novelty scores, such as Topics 1, 3, 8, and 11, involve a variety of fields such as data science, research evaluation, and patent technology and methodology, and have obvious interdisciplinary characteristics. These fields usually require the collaboration of researchers from several different backgrounds and fields to explore new research questions, methods, and applications, and the recombination of knowledge brought about by the collaboration of authors from different fields may be the cause of the recombination of knowledge in these topics. The recombination of knowledge brought about by the collaboration of authors from different fields may also be one of the reasons for the high novelty of papers in these topics.

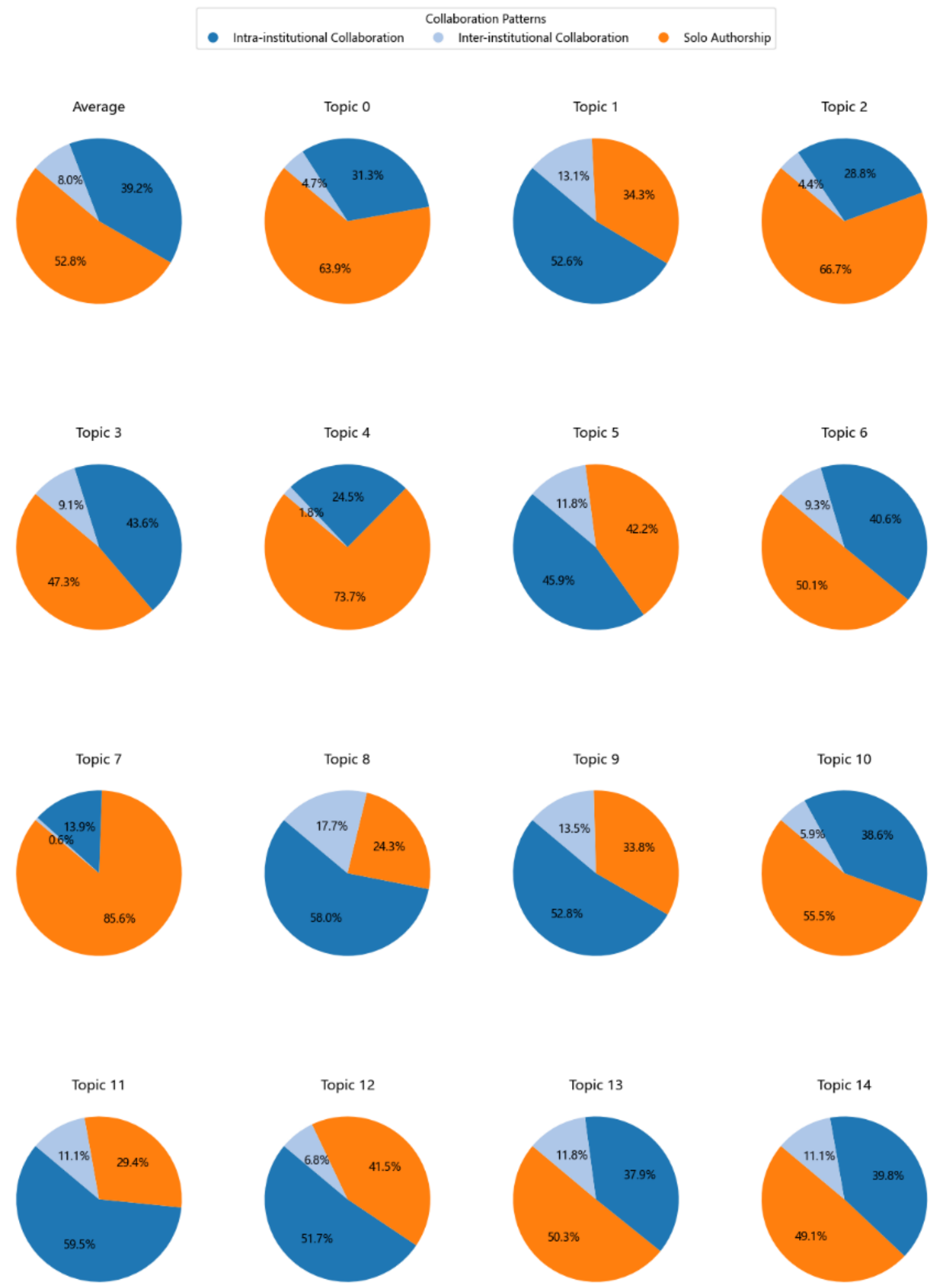

**Figure 12. Distribution of collaboration patterns in research papers on different topics**

# 5. Discussion

### *5.1 Implications*

This section discusses the theoretical and practical significance of this study.

The theoretical significance of analyzing the topic and novelty characteristics of papers in the field of LIS in China lies in: (1) In the process of topic categorization in this paper, by constructing a potential hierarchical tree of the topic distribution, we discover methodological connections and crossovers between different research fields, and promote integration and fusion between disciplines. (2) The results of this study's correlation analysis of author characteristics, such as the number of authors, and novelty scores support previous research that more authors from different institutions participating in scholarly creation contribute to a statistically higher level of novelty. (3) Meanwhile, this paper investigates the author collaboration patterns of academic papers with different novelties from the perspective of topic modeling, which can provide new perspectives and insights for theoretical studies such as correlation analysis of novelty in academic papers.

The practical significance lies in: (1) By analyzing and evolving the topics and novelties in the field of LIS in China, it can provide guidance and reference for the practical work of researchers, and help decision makers better understand the development trend and hot issues in different fields. (2) The formation and management of research teams can be optimized by analyzing the novelty scores of papers under different collaboration patterns. (3) Understanding the collaboration patterns under different topics can help formulate more effective academic policies. Policies for interdisciplinary collaboration can be promoted for topics with low collaboration intensity, promoting the integration and optimal use of academic resources.

### *5.2 Limitations*

However, this study has its limitations. Firstly, in this study, the BERTopic model was employed for topic modeling. Although based on the pre-trained BERT model, which captures richer semantic information of textual data, including word meanings, syntactic structures, contexts, etc., it is important to note that the BERT model is a deep neural network model lacking interpretability. Secondly, in calculating the novelty of papers, this study utilized the method proposed by LEE et al. based on the combinative innovation theory, which calculates novelty from a citation perspective. While leveraging a large amount of citation data enhances the scientific validity of the measurement results, it is important to note that this approach may deviate from the actual content of the articles compared to calculating novelty based on content. Lastly, due to the lack of standardization in early Chinese language corpora, some early research papers may lack references or abstracts. Consequently, these data had to be excluded from the research dataset, which could potentially impact the study’s conclusions. Finally, the level of paper novelty scores is the result of a combination of factors, and limited by the type of data available, this paper only examines the differences in paper novelty scores under different modes of collaboration from the perspective of statistical analysis.

# 6. Conclusion and Future Works

This study investigated the topic distribution of research papers in the field of LIS in China from 2000 to 2022, calculated the novelty of each paper, analyzed the distribution characteristics of novelty across different research topics, explored the evolutionary relationship of novelty over time, and finally analyzed the impact of collaboration pattern on novelty.

Firstly, this study conducted topic modeling on research paper abstracts in the field of LIS in China using the BERTopic model, resulting in 15 representative topics. The study introduced the topic characteristics and hierarchical structures of each topic, providing insights into the main research directions and hot topics within this field. The effectiveness of the topic model was validated through the calculation of topic coherence.

Secondly, this paper utilized the combinative innovation theory to calculate the novelty scores of each paper, and subsequently analyzed the distribution characteristics of novelty across different journals and topics. It was found that papers related to archival studies in journals such as Archives Science Bulletin and topics like topic13-Paper-File-Writing tended to have relatively lower novelty. Subsequently, empirical analysis was conducted by selecting papers with high and low novelty, and the distribution of topics in different novelty papers was analyzed. It was observed that papers with high novelty were mainly concentrated in topic3-Journals- Evaluation-Research and topic8 - Patents - Technology - Methods. Furthermore, the years were divided into intervals of 3 years to analyze the evolution of novelty in papers across different topics over time, revealing the developmental trends of novelty across different topics. Topic6-Culture-Public-Service and topic8-Patents-Technology-Methods exhibited the highest growth rates in novelty over the entire time range. Topic14-OA-OA Journal-Open had relatively high novelty scores in the early 21st century, but as the concept matured and became widespread, it experienced a declining trend in novelty scores from 2000 to 2006 due to a lack of new innovations and changes. However, with subsequent developments and changes, the novelty scores of this topic gradually recovered and increased. Overall, the novelty of research papers in the field of LIS in China showed a gradually increasing trend over time.

Thirdly, this paper examines and analyzes the novelty scores of papers in relation to the number of authors and the mode of co-authorship, in order to explore the differences in the mode of co-authorship of papers on different topics and its impact on the novelty of papers. Intuitively, papers with four co-authors tended to exhibit higher novelty, suggesting that multiple co-authors may come from different areas of specialization or disciplinary backgrounds, leading to a more comprehensive study and the discovery of more innovative points and novel content. In contrast, cross-institutional collaborations showed higher novelty compared to independent authorship and intra-institutional collaborative papers. This may be due to the fact that cross-institutional collaborations bring together researchers from different institutions, who bring diverse academic backgrounds, research methodologies, and perspectives, enough to look at the research problem from multiple perspectives and thus come up with more comprehensive and innovative solutions. In addition, analyzing the collaboration patterns of papers under different topics, we found that the proportion of independent authorships was significantly higher than the average for papers under topics with low novelty, while the proportion of collaborative papers tended to be higher for topics with high novelty scores, which may be due to the different degrees of interdisciplinarity involved in different topics or the different attention paid to novel research directions in different topics.

In the future, this study could be improved in the following three aspects: Firstly, exploring the use of different topic models to compare their effectiveness and obtain more objective and accurate topic classification results. Secondly, using novel metrics based on entities or topic vocabularies derived from the research content, rather than solely relying on citation-based methods for measuring novelty. This approach would enable a broader analysis that includes aspects such as the content of the papers. Lastly, regarding the analysis of collaboration pattern, this study only conducted superficial statistical analyses. Future research could delve into the causal analysis of the paper's novelty score and a number of aspects such as the number of authors, pattern of collaboration, financial support, academic age of the authors, and the degree of interdisciplinarity of the research topic.

## Acknowledgements

This study is supported by the Open Funding Project of Laboratory of ISTIC-Springer Nature Joint Laboratory for Open Science (Grant No. ISN23012) and the Postgraduate Research & Practice Innovation Program of Jiangsu Province. (Grant No. KYCX23_0601).